\documentclass[12pt,preprint]{aastex}

\def\AU{{\rm\, AU}}

\def\yr{{\rm\,yr}}

\def\pomega{\tilde{\omega}}

\begin{document}

\shortauthors{Chiang and Murray}
\shorttitle{Upsilon Andromedae}

\title{Eccentricity Excitation and Apsidal Resonance Capture in the Planetary
System Upsilon Andromedae}

\author{E.~I.~Chiang\altaffilmark{1} \& N. Murray\altaffilmark{2}}

\altaffiltext{1}{Center for Integrative Planetary Sciences,
Astronomy Department,
University of California at Berkeley,
Berkeley, CA~94720, USA}
\altaffiltext{2}{Canadian Institute for Theoretical Astrophysics,
60 St. George St., University of Toronto, Toronto, ON M5S 3H8, Canada}

\email{echiang@astron.berkeley.edu, murray@cita.utoronto.ca}

\begin{abstract}
The orbits of the outer two known planets orbiting Upsilon Andromedae
are remarkably eccentric. Planet C possesses an orbital eccentricity
of $e_1 = 0.253$. For the more distant planet D,
$e_2 = 0.308$. Previous dynamical
analyses strongly suggest that the two orbits are nearly
co-planar and are trapped in an apsidal resonance
in which $\Delta \pomega$, the difference between their
longitudes of periastron, undergoes a bounded oscillation
about $0\degr$. Here we
elucidate the origin of these large eccentricities
and of the apsidal alignment. Resonant interactions between
a remnant circumstellar disk of gas lying exterior
to the orbits of both planets can smoothly grow $e_2$.
Secular interactions between planets D and C can siphon off
the eccentricity of the former to grow that of the latter.
Externally amplifying $e_2$ during the phase of the apsidal
oscillation when $e_2/e_1$ is smallest drives the apsidal oscillation
amplitude towards zero. Thus, the substantial eccentricity
of planet C and the locking of orbital apsides are both consequences
of externally pumping the eccentricity of planet D over timescales
exceeding apsidal precession periods of order $10^4\yr$. We explain
why the recently detected stellar companion to $\upsilon$ And
is largely dynamically decoupled from the planetary system.
\end{abstract}

\keywords{planetary systems --- celestial mechanics --- stars: individual
($\upsilon$ Andromedae, HD168443, 47 UMa)}

\section{INTRODUCTION}
\label{intro}
The outer two known planets orbiting Upsilon Andromedae ($\upsilon$ And)
possess remarkably large orbital eccentricities (Butler et al.~1999).
Planet C
has a mass of $m_1 = (1.83 / \sin i_1) \, M_J$ (where $i_1$ is the angle
between the orbit pole of planet C and our line of sight, and
$M_J$ is the mass of Jupiter), an orbital
semi-major axis of $a_1 = 0.805$ AU, and an orbital eccentricity
of $e_1 = 0.253$. For planet D, the corresponding quantities
are $m_2 = ( 3.79 / \sin i_2) \, M_J$, $a_2 = 2.48 \AU$, and $e_2 = 0.308$.
These orbital parameters were kindly supplied by D. Fischer (2002,
personal communication), and represent more up-to-date values
than those employed by previous works.
Studies of the dynamical stability of the system (Rivera \& Lissauer 2000;
Stepinski, Malhotra, \& Black 2000; Lissauer \& Rivera 2001;
Chiang, Tabachnik, \& Tremaine 2001) strongly suggest that the two planets
execute approximately co-planar trajectories that are observed
nearly edge-on, so that $\sin i_1 \approx \sin i_2 \gtrsim 0.5$.

That $m_2$ likely exceeds $m_1$ while $e_2 > e_1$ stands at odds
with the idea that gravitational interactions between planets D
and C excited {\it both} $e_2$ and $e_1$ to their present-day values.
Close encounters between planets (Rasio \& Ford 1996; Weidenschilling \&
Marzari 1996; Ford, Havlickova, \& Rasio 2001) and mean-motion resonance
crossings between planets on divergent orbits (Chiang, Fischer, \& Thommes
2002) tend to impart the greater eccentricity to the less massive body.
Thus, we are led to the conclusion that an external agent---another planet,
a star, or the circumstellar disk from which the planets formed---must
have played a direct role in exciting the eccentricity of the most massive
planet, D. Of these three candidates, the latter is the most promising,
as we explain below.

\begin{figure}
\epsscale{1.0}
\plotone{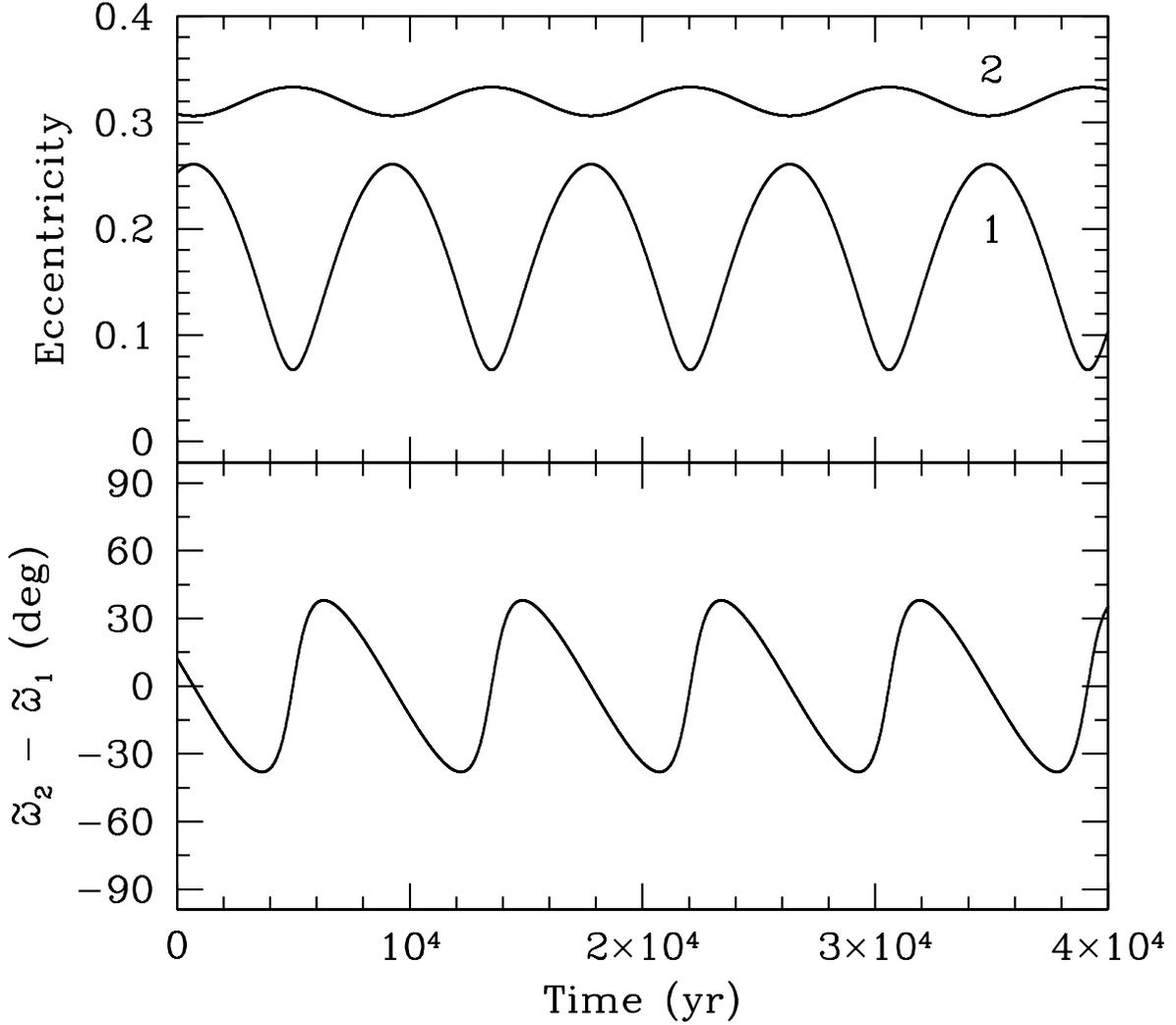}
\caption{Time evolution of orbital eccentricities of planets
C (denoted 1) and D (denoted 2), and the difference in their
apsidal longitudes, as computed using linear secular theory
for $\sin i_1 = \sin i_2 = 1$ and $i_{21} = 0\degr$.
The maximum in $\pomega_2 - \pomega_1$ is $38\degr$; this
differs from the preferred
value cited in CTT because of the inclusion of more recent
Doppler velocity measurements. The system spends
more time at small $e_2/e_1$ than at large $e_2/e_1$, a fact
that is important for understanding the damping of apsidal libration.
\label{laplag}}
\end{figure}

An additional clue to the origin of $\upsilon$ And lies in the apsidal
resonance that obtains if $i_{21}$, the mutual inclination between the orbits
of planets C and D, is less than $\sim$$20\degr$.\footnote{The
mutual inclination, $i_{21}$, does not necessarily equal $i_2 - i_1$;
see, e.g., Stepinski, Malhotra, \& Black (2000).}
The observation
today that the arguments of periastron of the two orbits, $\omega_1$
and $\omega_2$, differ by only $\sim$$12\degr$ has been used
by Chiang, Tabachnik, \& Tremaine (2001, hereafter CTT)
to argue that the orbits are indeed nearly co-planar.\footnote{The
argument of periastron, $\omega$, as fitted by Doppler velocity measurements is
referred to the plane of the sky (see, e.g., CTT).
The longitude of periastron, $\pomega$, may be referred to
any fixed line in inertial space for the purposes of this paper.
When $i_{21} = 0\degr$, we can take $\pomega = \omega$.} If
$i_{21} \lesssim 20\degr$, then $\Delta \omega \equiv \omega_2 - \omega_1$
librates about $0\degr$ with a semi-amplitude of $\sim$$38\degr$.
This apsidal libration and the eccentricity evolution of both planets
are displayed in Figure \ref{laplag}.

\placefigure{fig1}

This paper lays the groundwork for understanding the origin of
the large eccentricities and the apsidal alignment exhibited by the orbits
of planets C and D in $\upsilon$ And. Our main result is that the
eccentricity of planet C and the locking of orbital apsides
are both {\it consequences} of the slow
growth of the eccentricity of planet D.
The latter eccentricity, in turn, was driven by an external agent---plausibly
a primordial circumstellar disk lying exterior to the orbit of planet D---that
acted over timescales exceeding $10^4\yr$. We play our scenario out
and explain the mechanics of apsidal resonance capture in \S\ref{cap}.
We also discuss in that section why the recently detected
companion star to $\upsilon$ And
is not likely to be dynamically relevant. A summary is provided in \S\ref{sum}.

\section{Secular Growth of $e_1$ and Apsidal Resonance Capture}
\label{cap}

\subsection{Basic Model}

We take planets C and D to occupy co-planar, nearly circular
orbits at time $t = 0$. The dynamics exhibited by the system is qualitatively
the same over the range $0\degr \lesssim i_{21} \lesssim 30\degr$ (CTT).
We adopt current observed values for the orbital semi-major axes,
and minimum values for the planetary masses ($\sin i_1 = \sin i_2 = 1$).

We introduce an external force only on planet D that smoothly amplifies that
planet's eccentricity over a finite time interval.
The equations governing the eccentricities and apsidal longitudes ($\pomega$)
of the orbits of planets C and D read as follows (cf. Murray \& Dermott 1999):

\begin{eqnarray}
\label{ll}
\dot{h_1} & \equiv & \frac{d}{dt} (e_1 \sin \pomega_1) \, =  \, +A_{11} k_1 +
A_{12} k_2 \\
\label{ll2}
\dot{k_1} & \equiv & \frac{d}{dt} (e_1 \cos \pomega_1) \, =  \, -A_{11} h_1 -
A_{12} h_2 \\
\label{ll3}
\dot{h_2} & \equiv & \frac{d}{dt} (e_2 \sin \pomega_2) \, =  \, +A_{21} k_1 +
A_{22} k_2 + E h_2\\
\dot{k_2} & \equiv & \frac{d}{dt} (e_2 \cos \pomega_2) \, =  \, -A_{21} h_1 -
A_{22} h_2 + E k_2 \label{ll4}
\end{eqnarray}

\noindent where

\begin{eqnarray}
\label{a}
A_{11} & = & +\frac{n_1}{4} \frac{m_2}{m_{\ast} + m_1} \left( \frac{a_1}{a_2}
\right)^2 b^{(1)}_{3/2} \left( \frac{a_1}{a_2} \right) \\
A_{12} & = & -\frac{n_1}{4} \frac{m_2}{m_{\ast} + m_1} \left( \frac{a_1}{a_2}
\right)^2 b^{(2)}_{3/2} \left( \frac{a_1}{a_2} \right) \\
A_{21} & = & -\frac{n_2}{4} \frac{m_1}{m_{\ast} + m_2} \left( \frac{a_1}{a_2}
\right)^1 b^{(2)}_{3/2} \left( \frac{a_1}{a_2} \right) \\
\label{a4}
A_{22} & = & +\frac{n_2}{4} \frac{m_1}{m_{\ast} + m_2} \left( \frac{a_1}{a_2}
\right)^1 b^{(1)}_{3/2} \left( \frac{a_1}{a_2} \right) \\
E & = &\exp (-t/\tau_E) / \tau_e \, .
\end{eqnarray}

\noindent Here $n_j$ is the mean motion of planet $j$, $m_{\ast} = 1.3
M_{\odot}$ is the central stellar mass, $b^{(1)}_{3/2}$ and $b^{(2)}_{3/2}$ are
the usual Laplace coefficients,
and $1/E$ is the timescale over which exponential growth of $e_2$ occurs,
where $\tau_E$ and $\tau_e$ are fixed time constants.

We emphasize that only the eccentricity of planet D is directly amplified
by our external force. Our equations are intended to represent the following
physical picture, staged just after the formation epoch of the planets.
Interior to and between the orbits of planets C and D,
circumstellar disk gas is absent. Only exterior to the orbit of planet D
lies disk gas whose resonant interaction with that planet excites
the planet's eccentricity.
Our scenario is consonant with most current thinking on planet-disk
interactions. Gas between the two planets is plausibly
driven out of this region by planet-induced torques,
as numerical simulations by Kley (2000) and Bryden et al.~(2000)
find. Such gas either tunnels past both planets into the
innermost and outermost disks, or is accreted by the planets
(Bryden et al.~2000). Viscous accretion may drain away the disk
that lies interior to both planets onto the central star, as
has been invoked by Snellgrove, Papaloizou, \& Nelson (2001).
The disk outside $\sim$3 AU may be prevented from accreting inwards
by shepherding torques exerted by the outermost planet.
Goldreich \& Sari (2002) have discovered that disk gas can excite
a planet's eccentricity; eccentricity-amplifying Lindblad resonances
can defeat eccentricity-damping co-rotation resonances provided the
planet's initial eccentricity exceeds a critical threshold.
While the nonlinear outcome of their finite amplitude instability
remains to be worked out, it may be that a planet's
eccentricity ceases to grow when the planet's orbit begins to
overrun the gap edge, i.e., when $e \sim 0.3$.
For a dimensionless disk viscosity of $\alpha = 10^{-4}$ and
a disk mass of $40 M_J$, we compute an eccentricity
amplification timescale, $e_2/\dot{e_2}$, of $7 \times 10^4 \yr$, and
a radial migration timescale, $|a_2/\dot{a_2}|$, of $4 \times 10^6 \yr$
[Goldreich \& Sari 2002; their equations (18) and (19)].
Since the latter can exceed the former, we fix the semi-major
axes of both planets in our analysis.\footnote{We have verified
by direct numerical simulations that relaxing this assumption
changes none of our conclusions.}

We are aware that certain details underlying our scenario are
not well understood.
Gas between planets C and D might
be shepherded rather than driven out (Bryden et al.~2000),
in which case its effects on D and C would need to be
accounted for.
Numerical simulations by Papaloizou, Nelson, \& Masset (2000)
of interactions between Jupiter-mass planets and their parent disks
result in eccentricity damping and have
yet to be reconciled with the predictions of Goldreich \& Sari (2002).
Uncertainties in the plausibility of our scenario notwithstanding,
the key features of equations (\ref{ll})--(\ref{ll4}) are that
(a) the eccentricity evolution of planet D is directly driven by
an external influence, and (b) the eccentricity and apsidal evolution
of planet C are not, i.e., planet C feels only the gravitational
potential of planet D. These minimal requirements are sufficient
to understand the origin of the eccentricity of planet C and the origin
of the curious apsidal
alignment of the orbits of C and D, as we now demonstrate.

\begin{figure}
\epsscale{0.75}
\plotone{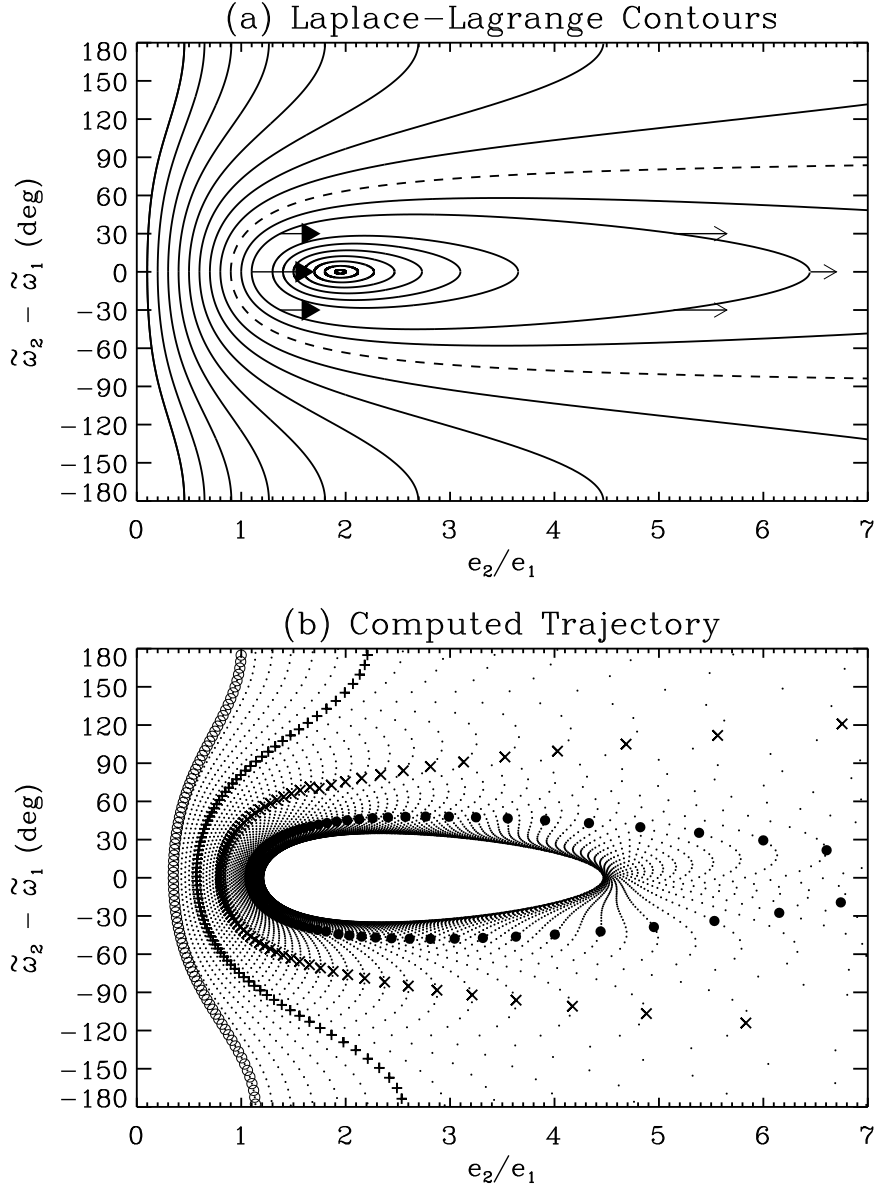}
\caption{(a) Solutions of equations (\ref{ll})--(\ref{ll4})
for $E = 0$. Each contour represents one possible solution.
The dashed contour closes at infinite $e_2/e_1$; it is the separatrix
dividing apsidally circulating solutions from apsidally librating ones.
(b) Solution of equations (\ref{ll})--(\ref{ll4}) for
$e_1 (t = 0) = e_2 (t=0) = 0.05$, $\Delta \pomega (t=0) = 180\degr$,
$\tau_e = 7.0 \times 10^4 \yr$, and $\tau_E = 1.5 \times 10^5\yr$,
over a time span of $1\times 10^6\yr$.
Each point marks the instantaneous position of the system in phase space.
Open circles denote the trajectory over a $8400\yr$ interval
starting at $t_s = 0$; $+$'s, $t_s = 5.1\times 10^4\yr$;
X's, $t_s = 1.2\times 10^5\yr$;
and filled circles, $t_s = 2.8\times 10^5\yr$. The system is captured from
circulation into libration and evolves towards smaller libration amplitude.
Damping of libration results from increasing $e_2$
when $e_2/e_1$ is smallest [filled arrows in (a)]. Increasing $e_2$
when $e_2/e_1$ is largest excites libration
[open arrows in (a)], but this effect is small
because the system spends less time at large $e_2/e_1$ than at small $e_2/e_1$.
\label{contour}}
\end{figure}

If $E = 0$, equations (\ref{ll})--(\ref{ll4}) yield
the classical Laplace-Lagrange (L-L) solution.
One solution, fitted with initial conditions appropriate
to observations of $\upsilon$ And today, is embodied in Figure 1.
A variety of L-L solutions, differing only in initial eccentricities
and apsidal longitudes, are mapped in the
space of $\Delta \pomega = \pomega_2 - \pomega_1$
and $e_2/e_1$ in Figure \ref{contour}a. The dashed
separatrix divides circulating solutions from
librating solutions. Librating solutions all have
$|\Delta \pomega| < 90\degr$.
Today planets C and D live on a contour of libration characterized
by $\max |\Delta \pomega| \approx 38\degr$.

\placefigure{fig2}

If $O(E) \ll O(A_{jk})$, or equivalently if $1/E$ is longer
than apsidal precession timescales of order $10^4 \yr$,
then the system evolves adiabatically through a series of L-L solutions.
Figure \ref{contour}b follows the evolution of a system for which
$e_1 (t = 0) = e_2 (t=0) = 0.05$, $\Delta \pomega (t=0) = 180\degr$,
$\tau_e = 7.0 \times 10^4 \yr$, and $\tau_E = 1.5 \times 10^5\yr$.
The final time of the calculation is $t_f = 1 \times 10^6\yr$,
at which time the eccentricity driving term, $E$, is effectively zero.
Each point represents the position of the system in
$\Delta \pomega$---$e_2/e_1$
space at a particular time; the solution was obtained by numerically
integrating equations (\ref{ll})--(\ref{ll4}) using a standard
fourth-order Runge-Kutta routine (Press et al.~1992). A set
of points highlighted by a symbol of a given type indicates the
trajectory over a time interval of $8400 \yr$.
Comparison of Figure \ref{contour}a
with \ref{contour}b indicates that the system morphs smoothly through
a series of L-L contours. What is noteworthy is that the system
is captured into libration and thereafter evolves towards
smaller libration amplitude.

The mechanics underlying apsidal resonance capture and the damping of
apsidal libration is as follows. The introduction of our eccentricity
driving term amplifies $e_2$, i.e., it pushes the system to the right in
Figure \ref{contour}.
Thus, a system initially occupying a contour of circulation is eventually
pushed past the separatrix onto a contour of libration. It
remains to explain the evolution towards smaller libration amplitude.
A librating system executes an elliptical trajectory in
$\Delta \pomega$---$e_2/e_1$ space. While the system traces out the left-hand
side of an elliptical contour,
external amplification of $e_2$ pushes the system
to the right, i.e., towards contours of smaller libration amplitude.
While the system traces out the right-hand side of the ellipse,
external amplification of $e_2$ also pushes the system to the right,
but now towards contours of larger libration amplitude. Which half of the
interaction dominates depends on how much time the system spends
on each half of the ellipse. It is clear from Figure \ref{laplag}
that the system spends more time near the minimum value of $e_2/e_1$
than near the maximum value of $e_2/e_1$.
The same behavior can be seen from Figure \ref{contour}b;
the points, plotted at equal time intervals, cluster
more strongly near small $e_2/e_1$ than near large $e_2/e_1$.
The reason for this is that the apsides of low
eccentricity orbits tend to precess more quickly than those of high
eccentricity orbits; for the same perturbative acceleration, $\dot{\pomega}
\propto \sqrt{1-e^2}/e$ by Gauss's equation; and thus when $e_1$ is minimal
($e_2/e_1$ maximal), the orbital apsides of planet C
precess more rapidly past the apsides of planet D
than when $e_1$ is maximal ($e_2/e_1$ minimal). Variations in $e_2$
can be ignored in this analysis; they are characteristically
smaller than variations in $e_1$ because planet D carries
the lion's share of the orbital angular momentum of the system
(see, e.g., CTT and Figure \ref{laplag}).
Thus, external driving of $e_2$ on the leftmost
half of an elliptical contour dominates the driving of $e_2$
on the rightmost half,
and the system spirals deeper into apsidal lock---all without
the need for any explicit dissipation of energy!

\begin{figure}
\epsscale{1.0}
\plotone{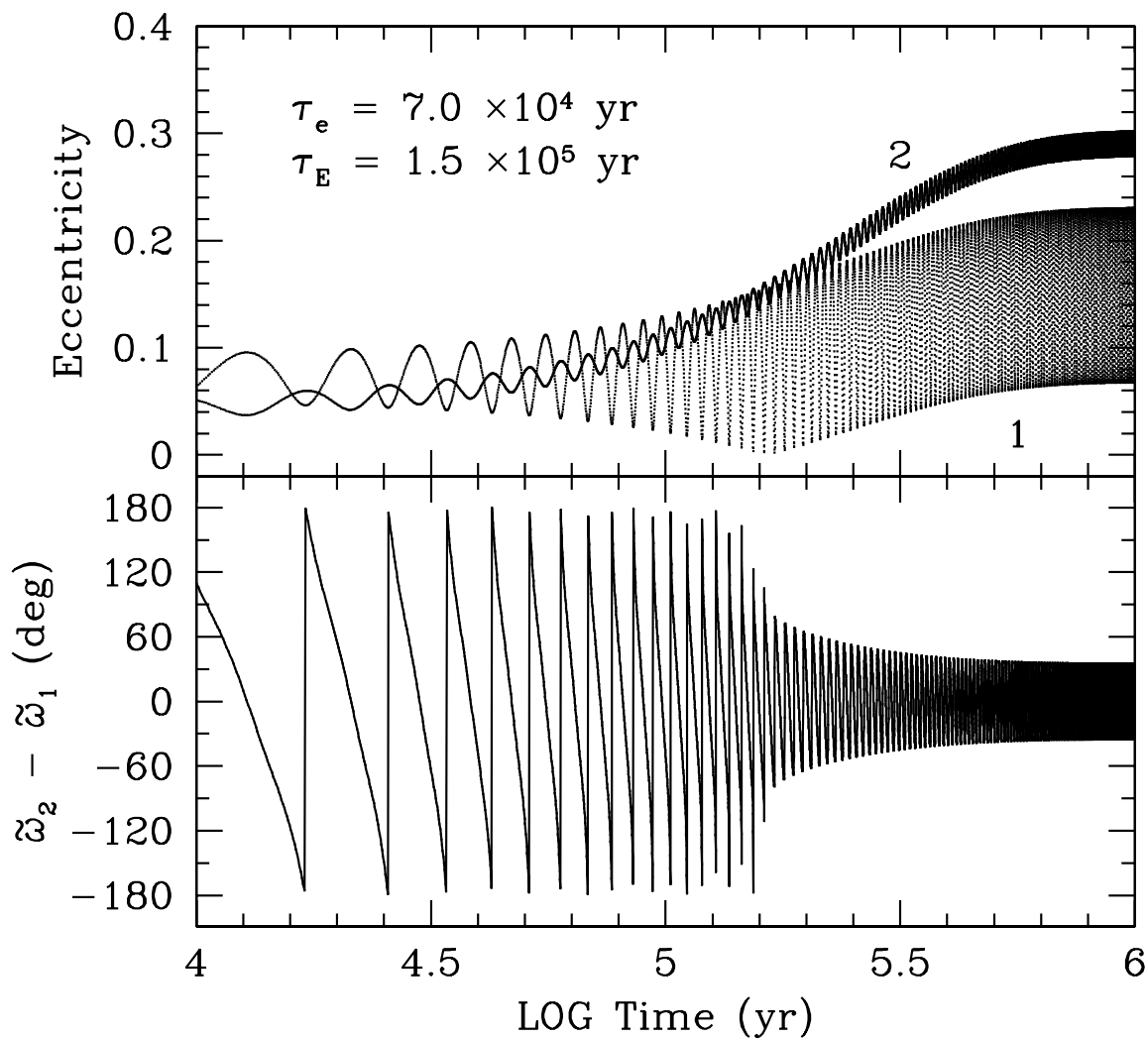}
\caption{Solution of equations (\ref{ll})--(\ref{ll4})
using input parameters listed under Figure \ref{contour}.
Externally driving the eccentricity of planet D amplifies
the eccentricity of planet C and locks the system into
apsidal resonance. At the end of the integration, the
eccentricities and apsidal longitude difference match
those of $\upsilon$ And today.
\label{time}}
\end{figure}

The celestial mechanical considerations described above are
general and we have verified by numerous integrations of equations
(\ref{ll})--(\ref{ll4}) that apsidal resonance capture and the subsequent
damping of apsidal libration occur under a wide variety of initial
conditions. The input parameters employed for the trajectory
shown in Figure \ref{contour}b were chosen so that the end state
of our calculation yielded system parameters that resemble those
of $\upsilon$ And today.
Figure \ref{time} portrays the same model, displaying the eccentricity
and apsidal evolution as explicit functions of time, and should
be compared with Figure \ref{laplag}.

\placefigure{fig3}

\subsection{Refinements and Extensions}

\subsubsection{Disk-induced Precession}

In addition to exciting the eccentricity of planet D,
an external disk would also cause the orbits of both planets
to precess at rates different from those due solely to the planets'
mutual gravity. We may model disk-induced precession by adding extra
terms to $A_{11}$ and $A_{22}$ that decay on timescales characterizing
the dissipation of the disk (by, e.g., photoevaporation).
This procedure can account for the disk's temporary
resonant and secular contributions
to planetary precession rates.
We have experimented by adding various transient terms
to $A_{11}$ and $A_{22}$ and find that none of our conclusions
changes qualitatively. Faced with extra, disk-driven differential precession,
we may still reproduce the observed parameters of $\upsilon$ And by
appropriately re-adjusting our input parameters (e.g., initial eccentricities
and $\tau_e$).

\subsubsection{Stellar Companion}

A stellar companion to $\upsilon$ And has recently
been reported by 2MASS (Two Micron All-Sky Survey; Lowrance, Kirkpatrick,
\& Beichman 2002). The mass of the companion is $m_E \sim 0.2 M_{\odot}$
and the projected separation between it and the primary is 750 AU.
Could this companion star be responsible for exciting the observed
eccentricities of planets C and D?

The answer is almost certainly no. The usual Kozai mechanism
for pumping planetary eccentricities (see, e.g., Holman, Touma,
\& Tremaine 1997) can only operate if (1)
the plane of the planetary system is sufficiently inclined
with respect to the plane of the stellar binary, and (2)
the apsidal precession rate of the planet due to the stellar companion
dominates other contributions to that precession rate. The
latter condition fails to be satisfied, by a wide margin.
We define $e_E$ and $a_E$ to be the (unknown) orbital eccentricity
and semi-major axis, respectively, of the companion star.
Then the precession period of planet D, as induced solely
by the stellar companion, is of order
$\sim 2\pi n_2^{-1} (m_{\ast}/m_E) (a_E/a_2)^3 (1-e_E)^{3/2}
\sim 2.5 \times 10^8 (0.2 M_{\odot}/m_E) (a_E / 750 \AU)^3
[(1-e_E) / 0.5]^{3/2} \yr$ (Holman, Touma, \& Tremaine 1997).
This greatly exceeds the precession
periods set by the mutual interaction between planets C and D,
which are of order $10^4 \yr$.

If the plane of the stellar binary has a small inclination
with respect to the plane of the planetary system, we may
treat the stellar companion as a third, massive body
in our linearized equations of motion. It is not difficult
to show that the companion star would then induce
an eccentricity of order $(a_2 / a_E) e_E$ in the orbit of
planet D. For the same parameters employed above, this secularly forced
eccentricity is of order $10^{-3}$.

Sigurdsson (1992) has shown that the mean induced
eccentricity and inclination of a planet orbiting
a $1.4 M_{\odot}$ star exceed $10^{-3}$ and $10^{-2}$,
respectively, if a $0.7 M_{\odot}$ companion approaches the primary
to within a distance of $\sim$5 times the semi-major axis
of the planet. These are mean induced orbital parameters,
averaged over 25,000 different possible encounter geometries.
For the stellar companion having $a_E \approx 750 \AU$
to come within $5 a_2 \sim 13 \AU$
of the primary, $e_E \gtrsim 0.98$. This is an underestimate
of the eccentricity required for the companion star to be dynamically relevant
because $m_E \approx 0.2 M_{\odot} < 0.7 M_{\odot}$.
In any case, we consider such large
eccentricities to be improbable. Moreover, if such large
eccentricities were to obtain, the consequent
violent forcing of the planetary system over the orbital period of the star
would threaten the stability of the system.

\subsubsection{Other Planetary Systems}

We are aware of two other extrasolar two-planet systems
whose present-day dynamics may be primarily secular: HD168443
(Marcy et al.~2001) and 47 UMa (Fischer et al.~2002).
In the first system, the less massive, inner planet
occupies the more eccentric orbit. This is consistent with the idea
that both planets excited each other's eccentricities by, e.g., divergent
resonance crossings (Chiang et al.~2002). Alternatively,
each planet's eccentricity may have been excited by local
disk material (Goldreich \& Sari 2002), independently of
the presence of the other planet. In HD168443,
the separation in semi-major axis between the two planets
is greater than that between planets C and D in $\upsilon$ And,
increasing the likelihood that disk gas was confined between the two planets
in the former system. In the absence of an innermost disk,
ring confinement is necessary for the resonant excitation of the
inner planet's eccentricity, either by divergent resonance crossings
or local disk-planet interaction. Both scenarios are consistent with
the fact that HD168443 does not exhibit an apsidal lock;
there is no reason to expect the ratio $e_{outer}/e_{inner}$
to be driven to large values.

In the case of 47 UMa, the eccentricity of the
less massive, outer planet is constrained
by observation to be less than 0.2 (Fischer et al.~2002).
If said eccentricity is less than $0.1$, we find that
the apsides of the two planets are locked about
$0\degr$, consistent with the findings of Fischer et al.~(2002).
Can we apply our theory for $\upsilon$ And to 47 UMa?
The eccentricities in 47 UMa are so low that it is not clear
how the system was driven by an external influence, if at all.
Moreover, the phase space of possible
trajectories in $\Delta \pomega$--$e_{outer}/e_{inner}$
space is qualitatively different from the case of $\upsilon$ And,
because most of the angular momentum of the system is carried by
the inner planet. This causes an island of libration about
$\Delta \pomega = 180\degr$ to appear at $e_{outer}/e_{inner} \gtrsim 1$.
An important feature of 47 UMa is the proximity of this
system to a 5:2 mean-motion resonance; if future Doppler velocity
data place the system in this resonance, then the apsides
circulate (Fischer et al.~2002). Whether or not the system inhabits
this mean-motion resonance, it seems that our theory for
$\upsilon$ And cannot be applied to 47 UMa in a straightforward manner.

\section{Summary}
\label{sum}

We have sketched an evolutionary scenario that explains the
origin of the eccentricities and the apsidal alignment
exhibited by the orbits of planets C and D in the extrasolar planetary
system, Upsilon Andromedae. In our picture, the eccentricity of
the outermost planet, D, grows smoothly via resonant interaction with
a circumstellar disk of gas remaining from the formation
of planets C and D. Provided the external, disk-driven amplification
of the eccentricity of planet D occurs over
a timescale that is long compared to apsidal precession
timescales set by the planets' mutual gravity, the orbits of planets C and D
evolve adiabatically through a series of classical Laplace-Lagrange
solutions.
Increasing the ratio of the eccentricity of planet D
to that of planet C eventually forces apsidal circulation
to give way to apsidal libration. Continued pumping
of this eccentricity ratio during the phase of the libration when
this ratio is smallest drives the libration amplitude to zero.
The resonantly excited eccentricity of planet D is secularly shared with that
of planet C. In short, the substantial eccentricity
of planet C and the apsidal resonance are both consequences
of the growth of the eccentricity of planet D which, in turn,
was driven by a third agent, most likely a remnant circumstellar disk
lying exterior to the orbits of both planets. The companion M4.5V
star to $\upsilon$ And recently detected by 2MASS is likely
dynamically decoupled from the planetary system.

Our mechanism for capture into apsidal resonance
and damping of apsidal libration is, to our knowledge,
novel. Moreover, it enjoys complete
independence from energy dissipation mechanisms that
are normally required to damp libration amplitudes as in, e.g.,
the case of the Laplace resonance inhabited by the Galilean satellites
(see, e.g., Murray \& Dermott 1999). The gravitational interactions
between planets C and D are, to excellent approximation,
secular and therefore energy-conserving.

\acknowledgements
It is a pleasure to acknowledge Geoff Marcy and Debra Fischer
for helpful and friendly exchanges and for providing
orbital parameters, and Patrick Lowrance and Chas Beichman
for generously sharing their most recent discovery.


\begin{references}

Bryden, G., Rozyczka, M., Lin, D.N.C., \& Bodenheimer, P.~2000, \apj, 540, 1091

Butler, R.P., et al. 1999, \apj, 526, 916

Chiang, E.I., Fischer, D., \& Thommes, E. 2002, \apjl, 564, L105

Chiang, E.I., Tabachnik, S., \& Tremaine, S. 2001, \aj, 122, 1607 (CTT)

Fischer, D. 2002, personal communication

Fischer, D., et al. 2002, \apj, 564, 1028

Ford, E.B., Havlickova, M., \& Rasio, F.A. 2001, Icarus, 150, 303

Gammie, C.F. 1996, \apj, 457, 355

Goldreich, P., \& Sari, R. 2002, \apj, submitted (astro-ph:0202462)

Holman, M., Touma, J., \& Tremaine, S. 1997, Nature, 386, 254

Kley, W. 2000, \mnras, 313, 47

Lissauer, J.J., \& Rivera, E.J. 2001, \apj, 554, 1141

Lowrance, P., Kirkpatrick, D., \& Beichman, C. 2002, \apjl, in press

Marcy, G., et al. 2001, \apj, 555, 418

Murray, C.D., \& Dermott, S.F. 1999, Solar System Dynamics (New
York: Cambridge University Press)

Papaloizou, J.C.B., Nelson, R.P., \& Masset, F. 2001, A\&A, 366, 263

Press, W.H., Teukolsky, S.A., Vetterling, W.T., \& Flannery, B.P. 1992,
Numerical Recipes in Fortran, Second Edition (New York: Cambridge University
Press)

Rasio, F., \& Ford, E.B. 1996, Science, 274, 954

Rivera, E.J., \& Lissauer, J.J. 2000, \apj, 530, 454

Sigurdsson, S. 1992, \apjl, 399, L95

Snellgrove, M.D., Papaloizou, J.C.B., \& Nelson, R.P. 2001, A\&A, 374, 1092

Stepinski, T.F., Malhotra, R., \& Black, D.C. 2000, \apj, 545, 1044

Weidenschilling, S.J., \& Marzari, F. 1996, Nature, 384, 619
\end{references}
\end{document}